\documentclass[11pt]{article}
\usepackage{amsmath,amssymb,color,graphics,epsfig}

\usepackage{amsfonts}

\newcommand{\hoch}[1]{$\, ^{#1}$}

\newcommand{\be}{\begin{equation}}
\newcommand{\ee}{\end{equation}}
\newcommand{\bea}{\setlength\arraycolsep{2pt} \begin{eqnarray}}
\newcommand{\eea}{\end{eqnarray}}

\def\ft#1#2{{\textstyle{\frac{\scriptstyle #1}{\scriptstyle #2} } }}
\def\fft#1#2{{\frac{#1}{#2}}}

\def\0{{\sst{(0)}}}
\def\1{{\sst{(1)}}}
\def\2{{\sst{(2)}}}
\def\3{{\sst{(3)}}}
\def\4{{\sst{(4)}}}
\def\5{{\sst{(5)}}}
\def\6{{\sst{(6)}}}
\def\7{{\sst{(7)}}}
\def\8{{\sst{(8)}}}
\def\sst#1{{\scriptscriptstyle #1}}
\def\oneone{\rlap 1\mkern4mu{\rm l}}

\begin{document}

\begin{flushright}
\end{flushright}

\vspace{25pt}
\begin{center}
{\large {\bf Exact Collapse Solutions in $D=4$, ${\cal N}=4$ Gauged Supergravity \\ and Their Generalizations}}

\vspace{10pt}
H. L\"u\hoch{1} and Xuefeng Zhang\hoch{2,1}

\vspace{10pt}

{\it \hoch{1}Department of Physics, \hoch{2}Department of Astronomy,\\
Beijing Normal University, Beijing 100875, China}

\vspace{40pt}

\underline{ABSTRACT}
\end{center}

We construct an exact time-dependent solution in $D=4$, ${\cal N}=4$ gauged supergravity, where the gauge fields of the $U(1)\times U(1)$ subgroup of the $SO(4)$ carry independent conserved charges. The solution describes a decaying white hole that settles down to the final state as a static charged black hole. We analyze the global structure and lift the solution back to $D=11$ supergravity. We further extend the theory by adding an extra term in the scalar potential and obtain a more general class of collapse solutions. The result constitutes a charged generalization of the Roberts solution and the dynamical scalar-hairy black hole solutions that have been very recently found by us.  The generalized Roberts solutions demonstrate that a scalar coupled to gravity can be unstable even when it is confined by a scalar potential with a fixed point.

\vfill {\footnotesize Emails: mrhonglu@gmail.com \ \ \ zhxf@bnu.edu.cn}

\thispagestyle{empty}

\pagebreak



\newpage

\section{Introduction}

The problem of gravitational collapse is of fundamental importance in general relativity \cite{Joshi07}. The study of black hole formation, as a principal outcome of such collapses, can shed light on a better understanding of spacetime singularities, cosmic censorship, critical phenomenon and gravitational waves (see recent reviews \cite{Joshi12,Gundlach07,Fryer11} and references therein). For matter sources that have been considered for collapse, two types of constructions exist. One is from certain generically specified matter energy-momentum tensor, with perhaps the most well-known example of Vaidya's metric and its various generalizations \cite{Vaidya51,Husain95,Wang99}. The other is from Lagrangians of Einstein gravity coupled with fundamental matter fields such as scalars or vectors (see, e.g., \cite{Roberts89,Guven96}). For the past few decades, much effort has been given to inspecting the collapse of a scalar field from both analytical and numerical aspects \cite{Gundlach07}. In particular, exact solutions from a Lagrangian are much harder to come by.

Parallel to this development of general relativity, many new solutions of charged static or stationary black holes have been constructed in supergravities. Supergravities are the low-energy effective theories of strings, which have been proposed as promising candidates for quantum gravity. However, very little is known regarding the dynamical process of continuous formation of these black holes. This paper is an attempt to fill this gap.

Recently, we have made some progress in finding exact (uncharged) solutions describing black hole formation in Einstein gravity minimally coupled to a scalar field with a specific self-interacting scalar potential \cite{zl}. The potential is supergravity-inspired, carrying two non-trivial parameters, the cosmological constant $\Lambda$ and an additional parameter $\alpha$. If $\alpha$ is turned off, we can embed the theory into $D=4$, ${\cal N}=4$ gauged supergravity.  Prior to our result, an example of \emph{charged} dynamical black hole solutions was first presented by G\"{u}ven and Y\"{o}r\"{u}k \cite{Guven96} in an Einstein-Maxwell-dilaton theory (without a scalar potential) that can be embedded in Poincar\'{e} supergravity. It is thus of interest to search for more non-static solutions that can arise in gauged supergravities.

In this paper, we consider the $U(1)\times U(1)$ truncation of the $SO(4)$ gauge fields of  $D=4$, ${\cal N}=4$ gauged supergravity \cite{Gates83} along with related generalizations. In section 2, we present a time-dependent solution with conserved $U(1)^2$ charges. We study its properties, particularly, the static limit as charged black holes. Furthermore, the solution is lifted back to $D=11$, and thereby represents time-evolving M2-branes with conserved angular momenta. In section 3, we push beyond the supergravity regime and include an extra $\alpha$-term in the scalar potential. A general class of charged solutions is derived, which contains as special cases both the supergravity solution and the time-dependent scalar-hairy black hole solutions found in \cite{zl}. We conclude the paper in section 4.

\section{Dynamic white/black hole formation}

\subsection{$D=4$, ${\cal N}=4$ (gauged) supergravity}

The bosonic sector of $D=4$, ${\cal N}=4$ supergravity consists of one metric, one complex scalar $\tau=\chi + {\rm i} e^{\phi}$ that forms an $SL(2,R)/U(1)$ coset and six abelian gauge fields, which, together with their Hodge duals, form six pairs of $SL(2,R)$-doublets. The theory can be gauged, giving rise to a scalar potential and $SO(4)\sim SU(2)\times SU(2)$ local gauged fields \cite{Gates83}.  Adopting the notation in \cite{Cvetic:1999au}, the Lagrangian for the bosonic sector is given by
\bea
{\cal L}_4 &=& R {*\oneone} - \ft12 {*d\phi}\wedge d\phi - \ft12 e^{2\phi} {*d\chi}\wedge d\chi + g^2 (4 + 2 \cosh\phi + \chi^2 e^{\phi}) {*\oneone}\cr
&&-\ft12 e^{-\phi} {*F^i \wedge F^i} - \ft12 \fft{e^\phi}{1 + \chi^2e^{2\phi}} {*\widetilde F^i \wedge \widetilde F^i}
-\ft12 F^i\wedge F^i + \ft12\fft{\chi e^{2\phi}}{1 +\chi^2 e^{2\phi}}
\widetilde F^i\wedge \widetilde F^i\,,
\eea
where $(F^i,\widetilde F^i)$ are the field strengths of the two $SU(2)$ gauge fields and ``$*$'' denotes the Hodge dual.  The parameter $g$ is the gauge coupling constant that yields an effective cosmological constant $\Lambda=-3g^2$ for AdS$_4$ vacuum solutions.

For our purpose, we focus on the $U(1)\times U(1)$ Cartan-subgroup truncation of the $SO(4)$ gauge group and the reduced Lagrangian reads
\be
e^{-1}{\cal L} = R - \ft12 (\partial \phi)^2 - \ft14 e^{\phi} F_1^2 - \ft14 e^{-\phi} F_2^2 + 2g^2 (2+\cosh\phi)\,,\label{lag}
\ee
with $F_i=dA_i$.  The truncation is consistent provided that
\be
F_i\wedge F_i=0\,,\qquad i=1,2.\label{cons}
\ee
This condition is required for setting the axion $\chi=0$. Note that the truncated theory can also be derived from the gauged STU supergravity model by setting the gauge fields pairwise equal. The equations of motion for (\ref{lag}) are
\bea
&&\Box \phi = \ft14e^{\phi} F_1^2 -\ft14 e^{-\phi} F_2^2+ \fft{dV}{d\phi}\,,\qquad \nabla_\mu (e^{ \phi} F_1^{\mu\nu}) = 0=\nabla_\mu (e^{-\phi} F_2^{\mu\nu})\,,\cr
&&E_{\mu\nu}\equiv R_{\mu\nu} - \ft12\partial_\mu\phi\partial_\nu \phi -
\ft12 V g_{\mu\nu} -\ft12e^{\phi} (F^2_{1\mu\nu} - \ft14 F_1^2 g_{\mu\nu})
-\ft12e^{-\phi} (F^2_{2\mu\nu} - \ft14 F_2^2 g_{\mu\nu})\,,
\eea
where $F_{i\mu\nu}^2 = F_{i\mu\rho}F_{i\nu}{}^\rho$ and $F_{i}^2 = F_{i\mu\nu}F_{i}^{\mu\nu}$, with the labeling index $i=1,2$.  Moreover, the theory can be further truncated consistently to Einstein-Maxwell gravity with the cosmological constant $\Lambda$ if one sets $F_1= F_2$ and $\phi=0$.

To construct time-dependent solutions, we start with the following ansatz in Eddington-Finkelstein-like coordinates:
\be
ds^2 = 2 du dr - H(r,u) du^2 + R(r,u)^2 d\Omega_{2,k}^2\,,
\ee
where
\be
d\Omega_{2,k}^2 = \fft{dx^2}{1 - k x^2} + (1-k x^2) dy^2\,,
\ee
with $k=(1,0,-1)$, is the metric for the unit 2-sphere $S^2$, 2-torus $T^2$ and hyperbolic 2-space $H^2$ respectively. The dilaton scalar $\phi$ is assumed to be a function of both $u$ and $r$, and the null coordinate $u$ is an advanced time which can be converted into a retarded time by a simple reversion $u\rightarrow -u$. In the paper, we will choose a time direction in which the dynamical solutions approach static black holes.  It follows that a specific solution may involve either the advanced or retarded time, depending on properties of the solution.

We consider the case when both $(F_1,F_2)$ carry only electric charges so that the constraint (\ref{cons}) is straightforwardly satisfied.  (The specific dilaton couplings of $F_i$ imply that solutions with only magnetic charges can be equally obtained by interchanging $F_1$ and $F_2$.) The equations of motion and the Bianchi identities of $F_i$ yield
\be
F_1 = \fft{4Q_1 e^{-\phi}}{R^2} du\wedge dr\,,\qquad
F_2 = \fft{4Q_2 e^{\phi}}{R^2} du \wedge dr\,,
\ee
where $Q_1$ and $Q_2$ are two conserved electric charges defined by
\be
Q_1 = \fft{1}{16\pi} \int e^{\phi} {*F_1}\,,\qquad
Q_2 = \fft{1}{16\pi} \int e^{-\phi} {*F_2}\,.
\ee
In this definition, we have implicitly assumed the principal orbits to be $S^2$, with the area $\omega_{2,k=1}=4\pi$.  For $k=0,-1$, we may also assume for simplicity that the principal orbits are some suitable Riemannian surfaces with $\omega_{2,k}=4\pi$.

The Einstein equation $E_{rr}=0$ gives us
\be
\fft{R''}{R} + \ft14 \phi'^2=0\,,
\ee
where a prime denotes the derivative with respect to $r$.  This equation is identical to its static counterpart except for the extra $u$-dependence. Hence we can solve it with known static solutions, but with the non-trivial integration constant replaced by a function of $u$. This leads to
\be
e^{\phi} = 1 + \fft{a(u)}{r}\,,\qquad R=\sqrt{r(r+a(u))}\,.
\ee
The unknown functions $H(r,u)$ and $a(u)$ can then be fully determined by the remaining equations of motion.  We will put forth a general class of solutions later in section \ref{section:gen}, when the theory is augmented by additional terms in the scalar potential.

\subsection{Time-dependent local solution} \label{section:sol1}

Our time-dependent charged solution in $D=4$, ${\cal N}=4$ gauged supergravity is given by
\bea
ds^2 &=& -\fft{2}{h^3} d\tilde u dr -  \fft{H}{h^6} d\tilde u^2 + r(r+a) d\Omega_{2,k}^2\,,\cr
A_1 &=& \fft{4Q_1}{(r+a)h^3} d\tilde u\,,\qquad A_2 = \fft{4Q_2}{r h^3} d\tilde u\,,\qquad
e^\phi =1 + \fft{a}{r}\,,\cr
H &=& g^2 r (r+a) + k  + \fft{8}{a}\Big(\fft{Q_2^2}{r} - \fft{Q_1^2}{r+a}\Big) +\fft{q}{2\tilde u_0} e^{-\tilde u/\tilde u_0}\,,\cr
a&=&\fft{q}{h}> 0\,,\qquad h(\tilde u)=\big(1 + e^{-\tilde u/\tilde u_0}
\big)^\fft12\,,\qquad \tilde u_0=\ft{q^3}{16(Q_1^2-Q_2^2)}>0\,.\label{sol1}
\eea
As before, $Q_i$'s are the two electric charges associated with the two field strengths. The solution assumes $Q_1> Q_2$. As we will see later, the apparent asymmetry between $Q_1$ and $Q_2$ in the metric is due to the choice of coordinates, and there is no loss of generality assuming $Q_1\ge Q_2$.  (The time-dependent $Q_1=Q_2$ solution can be obtained through some subtle limit and will be discussed later.)

To avoid having an implicit function in the metric (cf. section \ref{section:gen}), we have made a coordinate transformation $du=-d\tilde u/h^3$ with $h=q/a$.  In terms of the original null coordinate $u$, the function $a(u)$ satisfies
\be
a^3 \ddot a - 16 (Q_1^2 - Q_2^2) \dot a=0\,,
\ee
where a dot denotes the derivative with respect to $u$. More details on this equation will be given in section 3.

An important difference between the solution (\ref{sol1}) and the one we found in \cite{zl} is that the coordinate $\tilde u$ here is an retarded time, rather than an advanced time coordinate as was used in \cite{zl}. We have picked this time direction in order that the solution may have a physically reasonable end state ($\tilde u\rightarrow +\infty$) as a static black hole.

\subsection{The static limit}

Having acquired the time-dependent solution, we study the time evolution in details. First, we take the long-time limit $\tilde u\rightarrow +\infty$, for which we have $h\rightarrow 1$ exponentially fast with the characteristic relaxation time $\tilde u_0$. This results in an electrically-charged static black hole
\bea
ds^2 &=& -f dt^2 + \fft{dr^2}{f} + r(r + q) d\Omega_{2,k}^2\,, \qquad dt=d\tilde{u}+dr/f\,, \cr
A_1 &=& \fft{4Q_1}{r+q} dt\,,\qquad A_2 = \fft{4Q_2}{r} dt\,,\qquad
e^\phi =1 + \fft{q}{r}\,,\cr
f &=& g^2 r (r+q) + k  +\fft{8}{q}\Big(\fft{Q_2^2}{r} - \fft{Q_1^2}{r+q}\Big)\,,\label{sol2}
\eea
which has three independent parameters, i.e., two electric charges $Q_i$ and the parameter $q$ that is related to mass.  To examine the asymptotic structure at $r\rightarrow +\infty$, we introduce $\rho=\sqrt{r(r+q)}$.  At large $\rho$, we have
\bea
g_{tt} &=& - f = -g^2 \rho^2 - k + \fft{8(Q_1^2 - Q_2^2)}{q \rho} - \fft{4(Q_1^2 + Q_2^2)}{\rho^2} + \cdots\,,\cr
g^{-1}_{\rho\rho} &=& g^2\rho^2 + k + \ft14 g^2 q^2 - \fft{8(Q_1^2-Q_2^2)}{q\rho} +
\fft{4(Q_1^2 + Q_2^2) - \fft14 k q^2}{\rho^2} + \cdots\,.
\eea
It follows that the mass of the black hole is given by
\be
M_0=\fft{4}{q} (Q_1^2 - Q_2^2)\,.
\ee
Thus for a positive $q$ that has been assumed for our solution, we must impose $Q_1>Q_2$ for a positive mass.
The relaxation time can now be expressed as
\be
\tilde u_0 = \fft{4(Q_1^2 - Q_2^2)^2}{M_0^3}\,.\label{relax1}
\ee
This means that a bigger mass of the final black hole is accompanied by a faster decay, which is a feature that was also seen in \cite{zl}. The numerator here is somewhat intriguing and we will comment on its significance later.

When $q=2\sqrt2 (Q_1 - Q_2)$, we have
\be
M_0^{\rm BPS}=\sqrt2 (Q_1 + Q_2)\,,
\ee
and the solution (\ref{sol2}) becomes supersymmetric.  For ungauged supergravity corresponding to $g=0$, the BPS (supersymmetric) limit coincides with the extremal limit, in which the near-horizon geometry is AdS$_2\times S^2$.  The condition for a black hole to exist is $q\le 2\sqrt2 (Q_1 - Q_2)$ such that $M_0\ge M_0^{\rm BPS}$.  When this condition is satisfied, the horizon can be found at $r=r_0>0$ where $f(r_0)$ vanishes.  In gauged supergravity with $g\ne 0$, the BPS limit gives rise to superstar solutions with naked singularities.  The extremal limit requires that $M_0^{\rm ext}$ be greater than $M_0^{\rm BPS}$.

The Reissner-Nordstr\"om (RN) black hole cannot be retrieved from (\ref{sol2}) simply by setting $Q_1=Q_2$. A more involved limiting procedure is needed. We first reparametrize the solution as follows
\be
Q_1=\fft{Q}{\sqrt2} + \fft{mq}{8\sqrt2Q}\,,\qquad
Q_2=\fft{Q}{\sqrt2} - \fft{mq}{8\sqrt2Q}\,,
\ee
and then take $q\rightarrow 0$.  In this process, the dilaton vanishes and $Q_1=Q_2$, and the function $f$ turns into
\be
f=g^2r^2 + k - \fft{2m}{r} + \fft{4Q^2}{r^2}\,.
\ee
Thus with this choice of parameters, the conserved charge $Q_i$ has to take on an intricate relation with $q$ before one performs the limit. Since the constant $q$ becomes $a(u)$ in the time-dependent solution, while $Q_i$ remains conserved, our solution does not have a non-static limit from the Einstein-Maxwell theory with a vanishing scalar. This is consistent with the fact that the relaxation time vanishes when $Q_1=Q_2$, as given in (\ref{relax1}). This implies that the RN black hole in our truncated theory without additional matter fields may not be accessible from a dynamical process captured by our solution.

The asymmetrical way of $Q_1$ and $Q_2$ entering the metric is due to our choice of coordinates and parameters. Through the following coordinate transformation and reparametrization
\be
r\rightarrow r + \fft{m \sinh^2 (\delta_2)}{k}\,,\qquad
q=\fft{m}{k} \left( \sinh^2(\delta_1) - \sinh^2(\delta_2) \right)\,,\qquad
Q_i = \fft{m \sinh (2\delta_i)}{4\sqrt{2k}}\,,
\ee
the solution can be cast into
\bea
ds^2 &=& - (H_1H_2)^{-1} f dt^2 + H_1 H_2 \left(\fft{dr^2}{f} + r^2 d\Omega_{2,k}^2\right)\,,\qquad
e^{\phi} =\fft{H_1}{H_2}\,,\cr
A_1 &=& -\fft{\sqrt{2k}\cosh\delta_1}{H_1} dt\,,\qquad
A_2 = - \fft{\sqrt{2k}\cosh\delta_2}{H_2} dt\,,\cr
f &=& k - \fft{m}{r} + g^2 r^2 (H_1 H_2)^2\,,\qquad
H_i = 1 + \fft{m\sinh^2\delta_i}{k\,r}\,.
\eea
This form arises in the pairwise equal limit of the four-charged solutions in the gauged STU supergravity \cite{Duff:1999gh} for $k=1$ (see also \cite{tenauthor} for $k\ne 1$). One can now understand the apparent asymmetry of $(Q_1,Q_2)$ in the original coordinates.  For $Q_1^2<Q_2^2$, the positivity of the mass requires $q<0$.  Let $\tilde q=-q>0$, and redefine the coordinate $\tilde r = r-\tilde q$, and the solution becomes one with new parameters $\tilde Q_1\ge \tilde Q_2$ and $\tilde q>0$.  The transformation has a net effect of switching $A_1\leftrightarrow A_2$ and inverting $\phi\rightarrow -\phi$, which leave the Lagrangian invariant.  Therefore without loss of generality, one can always assume $Q_1\ge Q_2$ and $q>0$, which we shall do throughout the paper.

\subsection{Global structure and singularity}

The global structure of the static charged black hole is well understood. As mentioned earlier, we prefer a time direction such that a static black hole emerges at $\tilde u\rightarrow +\infty$. This can be done for our solution only when $\tilde u$ is a retarded time (as already chosen as such).  For the non-static solution, it is useful to adopt the luminosity coordinate $R$, defined as the radius of $d\Omega_{2,k}^2$.  In this radial coordinate, the metric reads
\bea
ds^2 &=&  - \fft{4 R\, d\tilde u dR}{h^2 \sqrt{q^2 + 4R^2 h^2}} - \fft{\tilde H}{h^6} d\tilde u^2 + R^2 d\Omega_{2,k}^2\,,\cr
\tilde H &=& g^2 R^2 + k + \fft{4(Q_1^2 + Q_2^2)}{R^2} -\fft{q^2(q^2 + 2 (h^2 +1) R^2)}{4\tilde u_0 R^2\sqrt{q^2 + 4 R^2 h^2}}\,.
\eea
For large $R$, we have
\be
\tilde H = g^2 R^2 + k - \fft{2M(\tilde u)}{R} + \fft{4(Q_1^2 + Q_2^2)}{R^2} + \cdots\,,
\ee
where the ``Vaidya mass'' $M(\tilde u)$ is
\be
M(\tilde u)= \fft{1 + \ft12 e^{-\fft{\tilde u}{\tilde u_0}}}{\sqrt{1 + e^{-\fft{\tilde u}{\tilde u_0}}}} M_0\,,
\qquad M_0=\fft{4}{q} (Q_1^2 - Q_2^2)\,.
\ee
As $\tilde u$ runs from $-\infty$ to $+\infty$, $M(\tilde u)$ monotonically decreases from $+\infty$ to a finite positive constant $M_0$. Therefore the solution can be interpreted as a white hole that radiates matter away while keeping the electric charges conserved till it settles down to the static limit that corresponds to a charged black hole.  The metric has a naked power-law curvature singularity at $R=0$ and an apparent horizon at $\tilde H(R,\tilde u)=0$ for some $R_0(\tilde u)>0$. As $\tilde u \rightarrow \infty$, the apparent horizon approaches the event horizon.  The analysis is analogous to that of \cite{zl} and we will omit the details.

\subsection{$Q_1=Q_2$} \label{section:solQ}

As we have commented earlier, for the time-dependent solution, one can choose without loss of generality $Q_1\ge Q_2$.  The solution (\ref{sol1}) was given for $Q_1>Q_2$. When $Q_1=Q_2$,  the solution (\ref{sol1}) appears to be static. In fact, we can take a special limit and still obtain a time-dependent solution (cf. section \ref{section:gen}); it is given by
\bea
ds^2 &=& 2 du dr - H du^2 + r (r + a) d\Omega_{2,k}^2\,,\cr
A_1 &=& \fft{4Q_1}{(r + a)} du\,,\qquad A_2 = \fft{4Q_1}{r} du\,,\qquad
e^\phi =1 + \fft{a}{r}\,,\cr
H &=& g^2 r(r+a) + k - q + \fft{8Q_1^2}{r(r+a)}\,,\qquad a=q u \,.\label{q1=q2}
\label{sol3}
\eea
Although one has two equal charges, the solution is not of Einstein-Maxwell gravity due to the non-vanishing of the dilaton. As we discussed before, the RN black hole cannot be obtained as a static limit of our time-evolving solution. Instead, the solution (\ref{sol3}) is a charged supergravity generalization of the Roberts solution \cite{Roberts89}.  The ``running'' dilaton in the Roberts solution is not unexpected since a free scalar is involved.  The situation is more puzzling in our case where the scalar potential has a stationary point at $\phi=0$. More specifically, the $u$-dependence of the dilaton in our solution is identical to that in the Roberts solution, regardless of the scalar potential. This suggests that the dilaton can still be unstable even under a confining scalar potential with a stationary point.

\subsection{Lifting the solution to $D=11$}

$D=4$, ${\cal N}=4$ gauged supergravity can be embedded in eleven-dimensional supergravity.  The explicit reduction ansatz for the bosonic sector was given in \cite{Cvetic:1999au} (see also \cite{tenauthor}). This allows us to obtain the corresponding metric in $D=11$:
\bea
ds_{11}^2 &=& \Delta^{\fft23}\Big[ds_4^2 + \fft{4}{g^2} \Big(d\xi^2 +
\fft{1}{r+a\,\cos^2\xi}\big( r\cos^2\xi\, d\Omega_3^2 + (r + a)\sin^2\xi\, d\widetilde \Omega_3^2\big)\Big)\Big]\,,\cr
d\Omega_3^2 &=& (d\psi + \cos\theta d\phi +\ft{1}{\sqrt2} g A_2)^2 + d\theta^2 + \sin^2\theta d\phi^2\,,\cr
d\widetilde \Omega_3^2 &=& (d\tilde\psi + \cos\tilde\theta d\tilde\phi+ \ft{1}{\sqrt2} g A_1)^2 + d\tilde\theta^2 + \sin^2\theta d\tilde\phi^2\,,
\eea
where the conformal factor is given by
\be
\Delta = \fft{r + a\,\cos^2\xi}{\sqrt{r(r+a)}}\,.
\ee
The solution depicts decaying, rotating M2-branes with two conserved angular momenta $(Q_1,Q_2)$ along the two $S^3$ transverse directions.

\section{Generalizations} \label{section:gen}

In this section, we discuss a more general theory by modifying the scalar potential with an additional term:
\be
e^{-1}{\cal L} = R - \ft12 (\partial \phi)^2 - \ft14 e^{\phi} F_1^2 - \ft14 e^{-\phi} F_2^2 + V(\phi)\,,
\ee
with
\be
V=-2g^2 (2+\cosh\phi) - 2\alpha^2 (2\phi + \phi \cosh\phi - 3 \sinh\phi)\,.
\label{pot}
\ee
The $\alpha$-term was first introduced in \cite{Zlosh05} for the pure scalar sector. It was later re-derived for the construction of neutral or charged scalar-hairy black holes \cite{Gonzalez13}, and also appeared in the time-dependent scalar-hairy black hole in \cite{zl}.

Adopting the same ansatz and following the same procedure as in section 2, we find that the theory admits the following time-dependent solution:
\bea
ds^2 &=&  2 du dr - H(r,u) du^2 + r\big(r + a\big) d\Omega_{2,k}^2\,,\cr
A_1&=& \fft{4Q_1}{r + a} du\,,\qquad A_2 = \fft{4Q_2}{r} du\,,\qquad
e^{\phi} = 1 + \fft{a}{r}\,,\cr
H&=& k + g^2 r (r + a) +\fft{8}{a}\Big(\fft{Q_2^2}{r} - \fft{Q_1^2}{r+a}\Big) - \dot a\cr
&&- \ft12 \alpha^2 a (2r + a) + \alpha^2 r\big(r+a\big)\log\big(1 + \fft{a}{r}\big)\,.\label{sol4}
\eea
where $a$ is a function of $u$ satisfying
\be
 \ddot a + \left(\alpha^2 a - \fft{16(Q_1^2 - Q_2^2)}{a^3}\right) \dot a=0\,. \label{cons1}
\ee
Here a dot denotes the derivative with respect to $u$.  It is worth pointing out that owing to the special dilaton coupling of the two field strengths, the contribution of the two electric charges to the Einstein and the scalar equations can be equivalently attributed to the electric and magnetic charges of the same field strength. The related solution is the same as (\ref{sol4}) with the constraint (\ref{cons1}) except that the field strengths are replaced by
\be
A_1= \fft{4Q_1}{r + a} du + 4Q_2 \omega_\1\,,\qquad d\omega_\1 = \Omega_{2,k}\,;\qquad A_2=0\,.
\ee
However it should be emphasized that this is no longer a supergravity solution even when $\alpha=0$, since the field strength $F_1$ does not satisfy (\ref{cons}).

The simplest solution of (\ref{cons1}) is $a$ being a constant ($a=q$), which corresponds to static charged black holes. The supergravity solution with $\alpha=0$ has been discussed in the previous section. Another noteworthy special case is when $Q_1=Q_2\equiv Q/\sqrt2$, and we have
\be
a=q \tanh(\ft12 \alpha^2 q\, u)\,.
\ee
(In fact, $a=q\coth(\ft12 \alpha^2 q\,u)$ and $a=-q\tan(\ft12 \alpha^2 q\,u)$ are also valid solutions of (\ref{cons1}), as one may check.) This function $a$ does not depend on $Q$ and was first obtained in \cite{zl} for the uncharged solution ($Q=0$). With the luminosity coordinate $R^2=r(r+a)$, the function $\tilde H$ (the coefficient of $du^2$) at large distance $R$ is given by
\be
\tilde H = g^2 R^2 + k - \fft{2M(u)}{R} + \fft{4Q^2}{R^2} + \cdots\,,
\ee
with the Vaidya mass
\be
M(u)=\ft1{12} a(3\dot a + \alpha^2 a^2) = \ft1{24} \alpha^2 q^2 \tanh(\ft12\alpha^2 q u) (3 - \tanh^2(\ft12\alpha^2 q u))\,,
\ee
which is independent of the electric charges.

We now move on to the general case of (\ref{cons1}). First we can calculate the Vaidya mass as
\be
M(u)= \ft14 a \dot a + \fft{4(Q_1^2-Q_2^2)}{a} + \ft1{12} \alpha^2 a^3\,.
\ee
It follows from the constraint (\ref{cons1}) that
\be
\dot M = \ft14 \dot a^2 \ge 0\,.
\ee
Thus in our solution, the mass increases with respect to the advanced time $u$ (or decreases with respect to the retarded time $v=-u)$.  This is consistent with the energy condition that is guaranteed in our Lagrangian formalism.

One way to solve the general case of (\ref{cons1}) is to note that it can be rewritten as
$\tilde h \ddot a + \dot a =0$ with
\be
\tilde h=\fft{a^3}{\alpha^2 a^4- 16(Q_1^2 - Q_2^2)}\,.
\ee
We can make a coordinate transformation $du=\tilde h d\tilde u$, so that the function $a$ can be solved analytically as follows:
\be
a(\tilde u) = \alpha^{-1} \sqrt{c+ e^{-\tilde u} \pm \sqrt{(c+e^{-\tilde u})^2 -16\alpha^2(Q_1^2 - Q_2^2)}}\,.\label{apm}
\ee
One may notice that for $\alpha=0$, the function $\tilde h$ is always negative, and hence the coordinate $\tilde u$ is a retarded time. Consequently the solution describes a decaying white hole as discussed before. If one has $Q_1=Q_2$ instead, $\tilde h$ is positive with $\tilde u$ being an advanced time, and then the solution captures gravitational collapse such as those in \cite{zl}. For a given solution with $\tilde u$ running smoothly from $-\infty$ to $+\infty$, the sign of $h$ remains the same. If initially $\alpha^4 a^4$ is sufficiently large, then $\tilde h$ is positive. With $\tilde u$ being the advanced time, the solution is of gravitational collapsing and $a$ grows bigger as $\tilde u$ increases, corresponding to the ``$+$''-sign of (\ref{apm}).  On the other hand, if initially $\alpha^4 a^4$ is small so that $\tilde h$ is negative, $\tilde u$ becomes the retarded time, in which case the solution is a white hole and $a$ turns smaller as $\tilde u$ increases, corresponding to the ``$-$''-sign of (\ref{apm}).

One may also keep using the null coordinate $u$ for solutions, but the drawback is that the function $a(u)$ can only be expressed as an implicit function of $u$ for the general range of parameters. To see this, we integrate the equation (\ref{cons1}) once and arrive at
\be
 {\dot a} = -\fft{L}{2a^2} - \fft{\alpha^2}{2} a^2 + c'\,, \qquad L = 16(Q_1^2-Q_2^2)\,,
\ee
where $c'$ is an integration constant. We can reparameterize this equation as
\be
 {\dot a} = -\fft{La^2}{2} \left(\fft{1}{a^2}-c_1^2\right)\left(\fft{1}{a^2}-c_2^2\right),
 \qquad c_1^2 c_2^2 = \fft{\alpha^2}{L},
 \qquad c_1^2 + c_2^2 = \fft{2c'}{L},
\ee
so that the general solution is given more conveniently by
\bea
 \fft{\mathrm{arctanh} (c_1 a)}{c_1} - \fft{\mathrm{arctanh} (c_2 a)}{c_2}
 + \ft12 L(c_1^2-c_2^2)u = 0, \qquad c_1^2 \neq c_2^2, \label{asol1} \\
 \fft{\mathrm{arctanh} (c_1 a)}{c_1} + \fft{a}{c_1^2a^2-1} - Lc_1^2u = 0,  \qquad c_1^2 = c_2^2. \label{asol2}
\eea
Here we have omitted the trivial translation $u\rightarrow u+c_0$. Depending on the values of $c'$, $\alpha^2$ and $L$, the two constants $c_{1,2}$ can become purely imaginary, in which case one may apply the identity
\be
 \mathrm{arctanh} (ix) = i\arctan(x)\,.
\ee
Moreover, changing any of the arctanh functions in (\ref{asol1}-\ref{asol2}) to arccoth still gives us valid solutions to (\ref{cons1}). This is due to the fact in the complex domain these two functions are only different by an additive complex number.

From the solution (\ref{asol1}), one can obtain the supergravity limit $\alpha^2=0$ by taking $c_1\rightarrow 0$. The result read
\bea
 a - \fft{1}{\sqrt{c}}\, \mathrm{arctanh}(\sqrt{c} a) = \ft12 Lcu , \qquad c>0, \label{asol3} \\
 a - \fft{1}{\sqrt{-c}}\, \arctan(\sqrt{-c} a) = \ft12 Lcu , \qquad c<0, \label{asol4}\\
 a = -\left(\ft32 L u\right)^{1/3}, \quad c=0, \label{asol5}
\eea
where $c=c_2^2$ is a real-valued integration constant. In the limit $c\rightarrow 0$, the first two cases above reduce to the last one by the relation $a - \fft{1}{\sqrt{c}}\, \mathrm{arctanh}(\sqrt{c} a) \sim -\ft13 ca^3$. The solution given in section \ref{section:sol1} corresponds to the case $c>0$. For the solution with $c<0$, we note that the linear term of $a$ will dominate over the $\arctan$ term for large $u$. Hence we have
\be
 a \sim \ft12 Lcu, \qquad u\rightarrow \infty.
\ee
This asymptotic linearity is quite similar to that of the Roberts solution \cite{Roberts89}. In fact, if one further takes the limit $L\rightarrow 0$ ($Q_1=Q_2$) and $c\rightarrow \infty$ while keeping the product $Lc=2q$ fixed, an exact linearity is acquired from (\ref{asol4}), i.e., $a = qu.$
This special case has been discussed in section \ref{section:solQ}.

To conclude this section, we point out that the time evolutions of $a(u)$ as given in (\ref{asol3}-\ref{asol5}) are in fact identical to those found by G\"{u}ven and Y\"{o}r\"{u}k (\cite{Guven96}, see eqns. (7.18-7.20)). However, their calculation is only for a single charge ($Q_2=0$) and without a scalar potential ($g^2=0=\alpha^2$).

\section{Conclusions}

In this paper, we have considered the $U(1)\times U(1)$ truncation of the $D=4$, ${\cal N}=4$ gauged supergravity and constructed two-charged time-dependent solutions.  We find that these solutions describe decaying white holes with constant charges and they eventually settle down to static and charged AdS black holes. The decaying process is exponentially fast with the relaxation time depending on the charges and the final black hole mass.  Our solution can be easily generalized to give rise to solutions in the $\omega$-deformed gauged STU supergravity \cite{Lu:2014fpa}.  Setting the gauge coupling constant to zero yields ungauged supergravity which contains the non-trivial global symmetry that allows one to generate the time-dependent dyonic solutions with the two field strengths all carrying both electric and magnetic charges.  This procedure however breaks down in gauged supergravities.  Nevertheless, dyonic generalization in gauged supergravities for static black holes were recently obtained \cite{Lu:2013ura,Chow:2013gba}.  We expect that this can be done for our time-dependent solutions as well.

We have also obtained the charged supergravity generalization of the Roberts solution.  In the landscape of string vacua, there can exist flat directions in the modulus space and they lead to instability that manifests in the Roberts solution.  Our solution (\ref{q1=q2}) shows that the dilaton can still be running even when it is confined by the potential with a stationary point.  Furthermore, the linear time dependence of the dilaton scalar charge in our solution remains the same as the Roberts solution, in spite of the extra scalar potential. Thus our construction may serve as a cautionary tale for the study of the modulus stability of a vacuum.

We further examine theories beyond supergravity by adding an additional $\alpha$-term in the scalar potential.  We obtain a more general class of charged time-dependent solutions that are asymptotically flat or (A)dS. It generalizes our recently found neutral black hole collapse solution \cite{zl}, as well as the supergravity solutions from section 2. Depending on the initial parameters, the general solutions can be decaying white holes or growing black holes, but all settle down to some final static states.

The low-energy effective theories of strings, namely (gauged) supergravities admit large classes of neutral or charged static and stationary black holes.  Unlike in general relativity where a black hole is a soliton of fundamental matter, some of these supergravity solutions describe the non-perturbative spectrum of the strings and M-theory. They play a crucial role in understanding the non-perturbative properties of the strings and M-theory. Our results may provide a first step of unlocking the time-dependent origin of these solutions.

\section*{Acknowledgement}

We are grateful to Zhong-Ying Fan, Hai-Shan Liu and Tong-Jie Zhang for useful discussions. The research is supported in part by the NSFC grants 11175269 and 11235003.

\end{document}